\documentstyle[12pt,epsfig]{article}

\columnwidth\textwidth

\begin{document}
\pagenumbering{arabic}
%-----------------------------
\everymath={\displaystyle}
\vspace*{-2mm}
\thispagestyle{empty}
\noindent
\hfill BUTP--96/15\\
\mbox{}
\hfill hep-ph/9607205\\
\mbox{}
\hfill  June 1996  \\
\vskip 3cm
\begin{center}
  \begin{Large}
  \begin{bf} {\Large \sc
Comparison of Lattice and Chiral Perturbation Theory calculations of
pion scattering lengths}
   \\
  \end{bf}
  \end{Large}
  \vspace{0.8cm}
   Gilberto Colangelo\\[2mm]
   {\em Institut f\"ur Theoretische Physik\\
    Universit\"at Bern\\
    CH-3012 Bern, Switzerland}\\[5mm]

\vskip 3cm

{\bf Abstract}
\end{center}
\noindent
I compare the lattice calculation of Fukugita et al. for the
pion $S$--wave scattering lengths to the predictions of Chiral
Perturbation Theory to two loop accuracy. I find good agreement,
despite the use of the quenched approximation in the lattice
calculation.


\newpage

In this letter I will present the comparison of two results which are
already available in the literature: the calculation of the pion
$S$--wave scattering lengths in quenched lattice QCD \cite{fuku} (see
also \cite{others}) and in Chiral Perturbation Theory (CHPT)
\cite{gl84} to two loops \cite{twoloop}.
Admittedly, the two results need not be the
same: the physical content of the two calculations is different since
QCD contains sea quark contributions, whereas the quenched lattice
calculation does not. Let me disregard this complication for the
moment, I will come to it later.

There is an important reason for attempting a comparison, which
however concerns more future developments than
present day calculations. As it is well known CHPT provides an
efficient technique for implementing the constraints of the chiral
symmetry of QCD on the Green functions. Remarkably, at low energy this
is enough to make predictions. This approach uses only the
symmetry properties of QCD: the dynamics of quarks and gluons is
totally ignored and shows up only in the values of some unknown low
energy constants, which we usually learn from experiments.
The major challenge for bridging the gap between QCD and its low
energy effective theory is to reproduce the value of these low energy
constants starting from the QCD Lagrangian. The best candidate for
doing this is lattice QCD, although very little has been
done up to know -- the only attempt in this direction of which I am
aware is the work by Myint and Rebbi \cite{myint}.

The comparison of lattice QCD and CHPT for quantities like the pion
$S$--wave scattering lengths is very interesting from this point of
view: finding agreement means that we understand the value of these
low energy constants -- as extracted from experiments -- on the basis
of the QCD Lagrangian. Of course the full accomplishment of this program
requires overcoming the quenched approximation.

The method to calculate scattering lengths on the lattice is due to
L\"uscher \cite{luescher}. In principle it allows to
extract these quantities in a theoretically very clean manner. The
method consists in measuring the energy of the fundamental state of
two static particles inside a spatial box of length $L$. L\"uscher has
calculated the expansion of this energy in inverse powers of the
length $L$, showing that the first three coefficients of this
expansion (which starts with a term of order $L^{-3}$) contain powers
of the $S$--wave scattering length. From the measurement of the energy
shift due to the finite size of the box one can extract the scattering
length. The method is very powerful, in particular because it allows
to directly relate a calculation  done in Euclidean space to a
quantity which is defined in Minkowski space. Moreover it transforms
what is usually considered a systematic error into an interesting
effect from which to extract useful information.

L\"uscher's formula can be applied only if $M_\pi L \gg 1$. This
condition was satisfied in the calculation of Ref.
\cite{fuku} by using an unphysical, moderately large pion mass. This
is not a problem for comparing to the CHPT calculation, since
in the chiral expansion the dependence on light quark masses is fully
explicit: given $\hat{m} = 1/2(m_u+m_d)$, or a value for $M_\pi$ and
$F_\pi$, one can calculate the corresponding value for the scattering
lengths. The only requirement for the
chiral expansion to be justified is that the pion mass be small with
respect to the typical mass scale of QCD, $M_\rho$ say. This condition
is satisfied in the case of staggered fermions,
where Fukugita et al. had $M_\pi/M_\rho \sim 0.3$. On the contrary, we
cannot compare to the Wilson fermions calculations, where
$M_\pi/M_\rho \sim 0.7$.

Let us discuss here in detail how the CHPT expressions have to be
calculated in the case of unphysical quark masses. For the purpose of
this discussion I will only consider the one loop case. The extension
to two loops is straightforward. At one one loop $a_0^0$ is given
by \cite{gl83}:
\begin{eqnarray}
\frac{32 \pi F_\pi^2}{M_\pi}a_0^0 &=& 7\left\{1+\frac{M_\pi^2}{F_\pi^2}
\left[
\frac{40}{7}\left(l_1^r(\mu)+l_2^r(\mu)\right)+\frac{10}{7}l_3^r(\mu)
+ 2 l_4^r(\mu) \right. \right.
\nonumber \\
 && \left. \left. - \frac{9}{32 \pi^2} \log
\frac{M_\pi^2}{\mu^2}   +\frac{5}{32\pi^2} \right] + O\left(
\frac{M_\pi^4}{F_\pi^4} \right)  \right\} \; .
\label{a00}
\end{eqnarray}
This expression is scale independent: the scale dependence of
the low energy constants $l_i^r(\mu)$ compensates the one due to the
chiral log.

To compare to the lattice result I use the ratio $M_\pi^2/F_\pi^2$ as
calculated on the lattice. The $l_i^r(\mu)$ do
not depend on the light quark masses and I use for them the values
given in Ref. \cite{twoloop}. The last thing to be evaluated is the
chiral log. This requires to give the value of $M_\pi$ in physical
units. To make the connection between lattice and physical units I
use the $\rho$ mass. This depends on the light quark masses
and can be expressed as a power series. Using an ansatz known as the
``quark counting rule'' (see {\em e.g.} Ref.\cite{quarkmasses}), I
consider only the first two terms in the expansion, according to:
\begin{equation}
M_\rho = M_V + 2 \hat{m} + O(\hat{m}^{3/2}) \; ,
\end{equation}
where $M_V$ stands for the $\rho$ mass in the chiral limit, for which
I use the value $M_V=760 \; \mbox{MeV}$. According to this ansatz, for
low values of $M_\pi/M_\rho$ the $M_\rho$ dependence on the light
quark masses makes a rather small effect, and the pion mass in
physical units, at $M_\pi/M_\rho =0.33$ turns out to be $M_\pi=260 \;
\mbox{MeV}$ (using also $B = 2 M_V$ \cite{quarkmasses}, where $B
\equiv -\langle 0 |\bar{q} q | 0 \rangle /F^2$).
It should be noted that the higher the value of the pion mass, the
higher is the relative importance of the low energy constants with
respect to the chiral log (at fixed $\mu$). This means that on the
lattice one can in principle increase the sensitivity of a given
quantity to the low energy constants, and improve the accuracy of
their determination.

The same procedure has to be repeated in the two loop case. There we
have six new constants appearing, the $r_i^r(\mu), \; \; i=1,\ldots 6$
whose value has been estimated via resonance saturation in Ref.
\cite{twoloop}. Here I use the same estimate. The effect of
these new constants is small.

\begin{figure}[t]
\epsfxsize 12 cm
\begin{picture}(30,1) \end{picture}
\epsffile{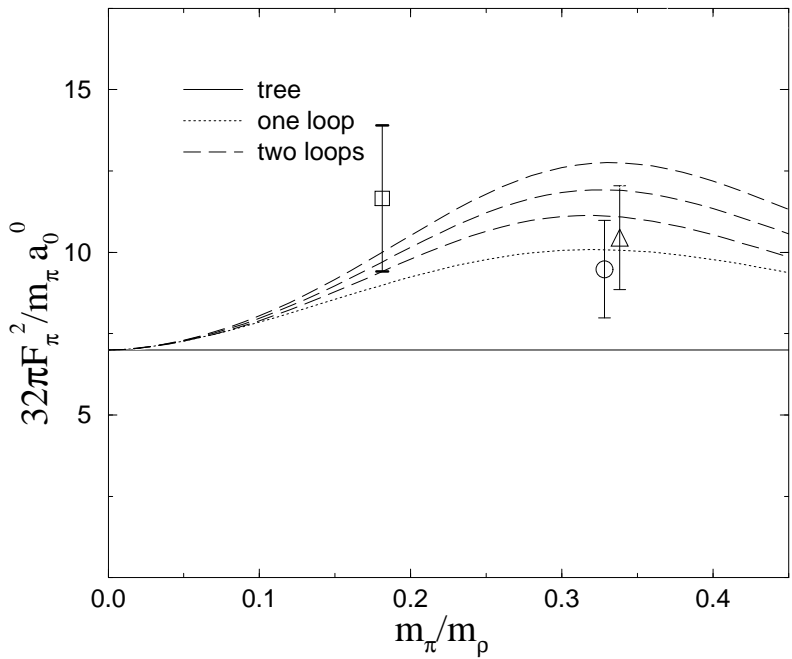}
\caption{Comparison of the $I=0$ pion scattering length as calculated
on the lattice by Fukugita et al. \protect{\cite{fuku}} and the CHPT
expansion. The data point with the square is the experimental measurement
\protect{\cite{rosselet}}. The two
lattice points are obtained with staggered fermions. The circle refers
to the calculation done without gauge fixing, and the triangle to the
Coulomb gauge.}
\label{fig1}
\end{figure}
%%%%%%%%%%%%%%%%%
\begin{figure}[t]
\epsfxsize 12 cm
\begin{picture}(30,1) \end{picture}
 \epsffile{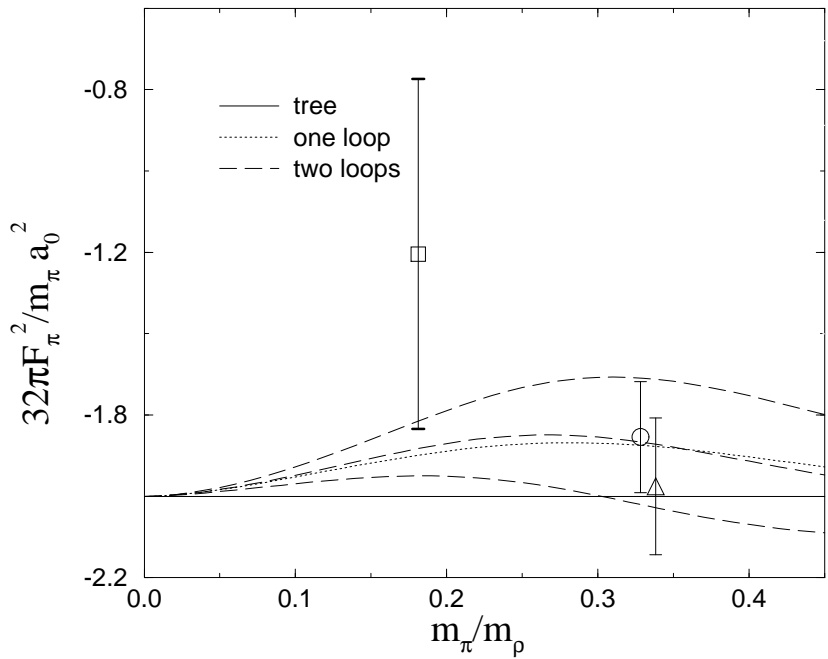}
\caption{Comparison of the $I=2$ pion scattering length as calculated
on the lattice by Fukugita et al. \protect{\cite{fuku}} and the CHPT
expansion. The legend for the data points is the same as in Fig. 1;
the experimental datum is from Ref. \protect{\cite{nagels}}}
\label{fig2}
\end{figure}
%%%%%%%%%%%%%%%%%%%%

The curves for the dimensionless quantities $32 F_\pi^2 a_0^I/M_\pi$
as functions of $M_\pi/M_\rho$ are shown in Figs. 1, 2 for $I=0,2$
respectively \cite{curves}.
The straight solid line corresponds to the Current Algebra ({\em i.e.}
CHPT to tree level) prediction \cite{weinberg},
whereas the dotted curve to the one loop CHPT calculation.
The three dashed curves are all calculated with the two loop
expression for the scattering length. The central one corresponds to
the values of the constants used in Ref. \cite{twoloop}, whereas the
other two are obtained by varying the low energy constant
$l_2^r(M_\rho)$ by $\pm 2 \; 10^{-3}$: this is a generous error
for this particular constant, and is meant to give a rough
estimate of the overall uncertainty as coming also from the other low
energy constants.

In Figs. 1 and 2 I have shown also the available experimental data
points \cite{rosselet,nagels}, and the lattice points calculated by
Fukugita et al. \cite{fuku}. In the $I=0$ case the two loop CHPT
curve is somewhat higher than the lattice points. In addition, if
one looks at the relative contribution of the two loops with respect
to the one loop and tree level, one can see that the series is not
converging very rapidly, and one may expect positive sizeable
contributions from higher orders.
In the $I=2$ case the one and two loop corrections to the
Current Algebra value are rather small, and the lattice calculation is
in agreement with all of them.
All in all, the lattice calculation and CHPT to two loops show
deviations from Current Algebra which agree in sign and size
surprisingly well, for both isospin cases.
This seems to indicate that the effects due to quenching are of the
same size, or smaller than the statistical errors of the lattice
calculations.

Finally, I come back to the question of a priori estimates of the
effect of quenching. In a recent article Bernard and Golterman \cite{bg},
have analyzed the modifications of L\"uscher's formula in the presence
of quenching. For this purpose they used the method called quenched CHPT
\cite{qCHPT}. In this approach one is able to obtain two types of
results: first, one can see whether in a specific quantity the
quenched approximation removes the chiral logs (in Ref. \cite{qCHPT}
they showed that in $M_\pi$ and $F_\pi$ to one loop there are none);
and secondly one can estimate the effects due to the $\eta$--singlet
propagator which in this approximation is ill--defined (because it
contains a double pole). These results show that the quenched
approximation introduces a qualitative change in approaching the
chiral limit. For finite quark masses, however, it is difficult to
estimate reliably the size of the change: in CHPT the chiral logs come
together with the low energy constants that remove the scale
dependence (see Eq. (\ref{a00})), and the splitting between the two is
arbitrary. At most one can try to estimate the part of the effect
which is due to the double pole in the $\eta$--singlet propagator (see
Ref. \cite{qCHPT} for details). This is what Bernard and Golterman
have done in the specific case of L\"uscher's formula \cite{bg}. Their
result is that the quenched approximation modifies considerably the
expansion of the energy shift in inverse powers of the box length: in
the $I=0$ case, for example, it introduces terms of order $1/L^n$ with
$n=0,2$. Nevertheless their numerical estimate of these spurious
effects is not too discouraging: for the staggered fermions
calculation of Fukugita et al. they estimate a $1\%$ effect in
the $I=0$ case (because of a strong cancellation), and a $20\%$
correction in the $I=2$ case.

This a priori estimate suggests that the quenched
approximation does not spoil completely the results of this lattice
calculation, as the comparison proposed here also indicates. On the
other hand, the use of quenching certainly diminishes the importance
of the agreement with two loop CHPT. In fact, had this result been
obtained in full QCD, one could conclude that lattice QCD is able to
explain the value of the two combinations of low energy constants
appearing in the two scattering lengths: this would be a very
remarkable achievement of lattice QCD.
Unfortunately, in the present case I cannot go that far, and I have to
conclude with the hope that there will be soon various improvements on
this lattice calculation.
For example, a repetition of this calculation with a different lattice
size would allow one to see whether the terms of order $1/L^n$ with
$n=0,2$ show up, and eventually to remove them. Moreover it would
allow a more reliable extraction of the coefficient of the $1/L^3$
term, by explicitly checking the volume dependence.
Hopefully, this and other future improvements will tell us
whether the agreement observed here contains some real physics, or is
only accidental.

\vskip 1 cm

{\bf Acknowledgements }
I thank for interesting discussions and/or support J. Gasser, P.
Hasenfratz, A. Papa, R. Petronzio and R. Urech , and J. Gasser for
very useful comments on the original manuscript.
This work has been supported in part by Schweizerischer Nationalfonds
and HCM--EEC--Contract No. CHRX--CT920026 (EURODA$\Phi$NE).

% ====================================================================


\begin{thebibliography}{99}
% ====================================================================

\bibitem{fuku}
M. Fukugita, Y. Kuramashi, M. Okawa, H. Mino, and A. Ukawa, Phys. Rev.
Lett. {\bf 71} 2387 (1993); Phys. Rev. D {\bf 52} 3003 (1995).

\bibitem{others}
M. Guagnelli, E. Marinari, and G. Parisi, Phys. Lett. B{\bf 240} 188
(1990);\\
S. R. Sharpe, R. Gupta, and G.W. Kilcup, Nucl. Phys. B{\bf 383} 309
(1992);\\
R. Gupta, A. Patel, and S. R. Sharpe, Phys. Rev. D {\bf 48} 388
(1993).

\bibitem{gl84}
J. Gasser, and H. Leutwyler, Ann. of Phys. (N.Y.) {\bf 158} 142 (1984).

\bibitem{twoloop}
J. Bijnens, G. Colangelo, G. Ecker, J. Gasser, and M. Sainio, Phys.
Lett. B {\bf 374} 210 (1996).

\bibitem{myint}
S. Myint, and C. Rebbi, Nucl.Phys. B{\bf 421} 241 (1994).

\bibitem{luescher}
M. L\"uscher, Comm. Math. Phys. {\bf 105} 153 (1986).

\bibitem{gl83}
J. Gasser, and H. Leutwyler, Phys. Lett. {\bf 125}B 325 (1983).

\bibitem{quarkmasses}
J. Gasser, and H. Leutwyler, Phys. Rep. C{\bf 87} 77 (1982).

\bibitem{rosselet}
L. Rosselet et al., Phys. Rev. D {\bf 15} 574 (1977).

\bibitem{nagels}
M.M. Nagels et al., Nucl. Phys. B{\bf 147} 189 (1979).

\bibitem{curves}
To obtain the curves I had to vary the ratio
$M_\pi/F_\pi$ continuously, letting it go through 2.24 at
$M_\pi/M_\rho=0.18$ (physical value), and 4.8 at $M_\pi/M_\rho=0.33$
(lattice value, see Ref. \cite{fuku}).
How I interpolated between these values is irrelevant for the present
discussion.

\bibitem{weinberg}
S. Weinberg, Phys. Rev. Lett. {\bf 17} 616 (1966).

\bibitem{bg}
C.W.  Bernard, and M.F.L. Golterman, Phys. Rev. D {\bf 53} 476 (1996).

\bibitem{qCHPT}
C.W. Bernard, and M.F.L. Golterman, Phys. Rev. D {\bf 46} 853 (1992).

\end{thebibliography}
\end{document}